\documentclass[12pt]{article}
\usepackage{lmodern}
\usepackage{newtxtext,newtxmath}
\usepackage{graphicx}
\usepackage{makecell}
\usepackage[letterpaper,margin=1in]{geometry}

\renewenvironment{abstract}
	{\quotation}
	{\endquotation}

\date{}

\makeatletter
\renewcommand{\fnum@figure}{\textbf{Figure \thefigure}}
\renewcommand{\fnum@table}{\textbf{Table \thetable}}
\makeatother

\usepackage{scicite}

\usepackage{url}

\def\scititle{
	Meta-Optics-Based Diffractive Learning Machine for Angle-Multiplexed Multi-Task Processing
}
\title{\bfseries \boldmath \scititle}

\author{
    Mingcheng~Luo$^1$,
    Jianmin~Xiong$^1$,
    Jiayong~Peng$^1$,
    Jingtian~Hu$^2$,
    Hongwei~Chen$^3$,\\
    Chester~Shu$^1$,
    Chaoran~Huang$^{1\ast}$\and
	\small$^{1}$Department of Electronic Engineering, The Chinese University of Hong Kong, Shatin, New Territories, Hong Kong\and\
	\small$^{2}$Ministry of Industry and Information Technology Key Lab of Micro- Nano Optoelectronic Information System,\and 
    \small Guangdong Provincial Key Laboratory of Semiconductor Optoelectronic Materials and Intelligent Photonic Systems, \and
    \small Harbin Institute of Technology, Shenzhen 518055, China.\and
	\small$^{3}$Department of Electronic Engineering, Tsinghua University, Beijing, China. \and
	\small$^\ast$Corresponding author. Email: crhuang@ee.cuhk.edu.hk
}

\begin{document} 

% Insert the title and author list
\maketitle

% Abstract, in bold
% There are strict length limits, and not all formats have abstracts.
% Consult the journal instructions to authors for details.
% Do not cite any references in the abstract.
\begin{abstract} \bfseries \boldmath
% Start with one or two sentences of background
All-optical diffractive neural networks (DNNs) offer low latency, low energy consumption, and massive parallelism. However, their scalability in terms of network scale and multi-task processing remains limited. Here, we experimentally demonstrate a large-scale metasurface-based diffractive learning machine for angle-multiplexed multi-task processing. Our system employs a metasurface-based reflection cavity to realize a five-layer DNN comprising 115 million fixed, untrained diffractive nodes within a compact volume of 216 mm$^3$. By integrating this large-scale, untrained DNN with a lightweight, trainable digital backend containing only 7,680 parameters, our system experimentally achieves an accuracy of 95.3\% on the standard full-scale CIFAR-10 dataset, significantly outperforming previous optical neural networks and rivaling large-scale digital models. Using a novel incident-angle-multiplexing scheme, our system experimentally achieves parallel processing of 11 tasks with negligible performance degradation, representing the largest multi-task processing capacity reported to date. Our work provides a scalable and energy-efficient solution for high-performance edge computing.
\end{abstract}

% The first paragraph of any Science paper does NOT have a heading
% Nor is it indented
\section{Introduction}
\noindent
Artificial intelligence (AI) has been reshaping our daily lives by boosting a range of applications, such as autonomous driving, human-like robots, and smart healthcare~\cite{ma2020artificial,lee2021application,jacobsen2004research}. The heart of AI lies in artificial neural networks (ANNs). In the past decade, ANNs have been continuously growing in model scale, so as to tackle intricate tasks, such as natural language processing, image recognition, and autonomous decision-making~\cite{min2023recent,hacker2023regulating}. As a result, substantial computational resources are required to implement ANNs, which poses a challenge in the development of computing hardware. Conventional electronic computers have achieved remarkable computing performance; however, the performance improvements of electronic computers have progressed slowly in the post-Moore era. Consequently, it remains a key challenge to meet the rapidly escalating computational demand for AI.

All-optical diffractive neural networks (DNNs) have been regarded as a promising alternative to ANNs, due to the ability to process optical information at the speed of light in a large-scale parallel way\cite{lin2018all,zhou2021large,luo2025large,duan2023optical,luo2022metasurface,zhang2023advanced,luo2019design,zheng2022meta,zheng2024multichannel,li2019intelligent,goi2022direct,luo2024meta,luo2026highly,peng2026optical}. DNNs typically consist of multiple cascaded two-dimensional (2D) diffractive layers, each of which can provide high-density, all-to-all diffractive nodes. With the availability of sufficient diffractive nodes, DNNs can provide rich transformation space solutions. Flat optical metasurfaces are appealing building blocks of DNNs. Optical metasurfaces consist of a two-dimensional array of sub-wavelength-scale optical resonators (known as meta-atoms)~\cite{chen2016review,hsiao2017fundamentals,wang2017broadband,yang2020all,arbabi2017planar}. Unlike traditional diffractive optical elements (DOEs) that can only modulate optical phase, optical metasurfaces can be designed to manipulate the phase, amplitude, and polarization of the light~\cite{overvig2019dielectric,balthasar2017metasurface}. This unique feature empowers metasurface-based DNNs to realize a variety of functionalities that DOE-based DNNs cannot achieve~\cite{luo2025large,wang2024matrix,luo2022metasurface,zheng2022meta,zheng2024multichannel,luo2024meta}. For example, recent demonstrations of metasurface-based DNNs have achieved multi-task processing using polarization multiplexing and optical convolution using polarization conversions~\cite{luo2022metasurface,zheng2022meta}. Furthermore, the size of meta-atoms is sub-wavelength-scaled, which is much smaller than that of the diffractive units of conventional free-space optical devices, offering metasurface-based DNNs with compactness and miniaturization advantages~\cite{zheng2024multichannel,faraji2018compact,kim2024metasurface}.

However, metasurface-based DNNs still face challenges in increasing the number of diffractive layers and nodes, which causes performance gaps with digital competitors. These limitations stem from several key factors. First, training DNNs demands high-precision computation to model their physical processes numerically, a task that becomes exceptionally difficult as both the number of layers and the total parameters scale upward~\cite{zhou2020situ, zheng2023dual}. Moreover, even after training, DNNs inevitably suffer both fabrication errors and physical implementation errors\cite{mengu2020misalignment}, ultimately degrading performance. The above issues are further intensified as the number of parallel tasks increases in multi-task processing, where tasks are encoded using different light properties (for example, wavelengths and polarizations). This requires the joint optimization of multiple physical models, leading to competition among tasks and overall performance degradation as the number of parallel tasks grows.~\cite{duan2023optical,liu2025ultra}. As a consequence, current DNNs can only experimentally implement at most three tasks in parallel~\cite{duan2023optical,luo2022metasurface,wang2024matrix,chi2025metasurface,wang2025high,liu2025polarization,liu2025ultra}. 

To solve the challenges of training large-scale DNNs and ensure their generalizability and scalability to multi-task learning, we leverage the concept of the Extreme Learning Machine (ELM)~\cite{huang2006extreme,huang2015trends}. This approach uses randomized and fixed hidden-layer weights, with only the output layer being trained. By training only this lightweight output layer, the model becomes highly adaptable to new tasks, as the fixed hidden layer provides a universal feature space transformation. This training-free expansion of hidden layers eliminates both the training burden and fabrication errors typically associated with scaling depth and parameter counts. Consequently, the network can be extended substantially without the performance degradation seen in conventional DNNs, leading to significantly enhanced overall performance and multi-tasking capability, as demonstrated by the experimental results presented in this work.

By adopting this strategy, we experimentally demonstrate a large-scale metasurface-based diffractive learning machine (Meta-DLM) for angle-multiplexed multi-task vision processing. Our Meta-DLM comprises a metasurface-based five-layer all-optical DNN containing 115 million fixed, untrained diffractive nodes, coupled with a lightweight, trainable digital backend. Crucially, the optical frontend employs a reflective-cavity architecture that scales the effective network depth simply by increasing the number of light round trips, while maintaining an ultra-compact volume of only 216 mm$^3$. This design achieves a substantially smaller form factor than previously reported reflective-cavity-based learning machines~\cite{xia2024nonlinear,yildirim2024nonlinear,dong2025scalable}. Experimentally, our system achieves an accuracy of 95.3\% on the standard full-scale CIFAR-10 dataset using only 7,680 trainable digital parameters, substantially outperforming previous optical neural networks and rivaling large-scale digital models with millions of trainable parameters~\cite{real2019regularized,hu2018squeeze,cubuk2019autoaugment}. Moreover, we leverage incident-angle multiplexing to enable multitask operation, whereby distinct tasks are naturally multiplexed and demultiplexed according to the input angle, eliminating the need for additional wavelength- or polarization-demultiplexing components. Using this approach, we experimentally scale the system to 11 parallel tasks, representing, to the best of our knowledge, the largest multitask processing capacity reported to date~\cite{duan2023optical,luo2022metasurface,wang2024matrix,chi2025metasurface,wang2025high,liu2025polarization,liu2025ultra}. Together, these advances establish Meta-DLM as a compact, large-scale edge-computing platform for machine vision with unprecedented multi-task processing capability.

\section{Results}
\subsection{System description and working principle}\label{sec2.1}

\textcolor{blue}{Figure 1a} illustrates the implementation of the proposed Meta-DLM for angle-multiplexed multi-task processing. Multi-task operation is achieved by encoding each task into a distinct incident angle: digital inputs from $N$ tasks are converted into optical images using spatial light modulators (SLMs) and then projected into the double-layer metasurface-based reflection cavity at angles $\theta_1, \theta_2, \ldots, \theta_N$. Each metasurface comprises $4{,}800 \times 4{,}800$ (approximately 23 million) cylindrical silicon meta-atoms with a fixed period of 500 nm and diameters ranging from 100 to 400 nm. At the operating wavelength of 532 nm, each meta-atom functions as a reflective element whose complex reflection coefficient is determined by the diameter of the corresponding meta-atom~\cite{hsiao2017fundamentals}. Here, the phase-modulation coefficients of the meta-atoms are randomly selected from a prescribed Gaussian distribution, whereas the amplitude-modulation coefficients are not independently designed. These complex-valued coefficients remain untrained and become fixed after fabrication. Further details of the metasurface design and fabrication are provided in \textcolor{blue}{Methods} and \textcolor{blue}{Supplementary Note 1}.

Within this cavity, the input field ($\mathbf{U}_0$) undergoes five round-trip reflections before readout. During each reflection, the field interacts with a distinct spatial region of the metasurfaces and thus experiences a different modulation matrix with entries $(w_1^i,\ldots,w_m^i)$, where $w_m^i$ denotes the complex-valued modulation coefficient of the $m$-th meta-atom during the $i$-th reflection. Consequently, the output field can be expressed as $\boldsymbol{M}\mathbf{U}_0$, where the transfer matrix ($\boldsymbol{M}$) is given by

\begin{equation}
    \boldsymbol{M} = \prod \limits_{i=1}^5 (\boldsymbol{W}_ddiag(w_1^i, \dots w_m^i))
\end{equation}
\noindent where is $\boldsymbol{W}_d$ denotes an $m \times m$ matrix characterizing free-space diffraction and $diag(w_1^i, \dots, w_m^i)$ denotes an $m \times m$ diagonal matrix whose diagonal entries are $(w_1^i, \dots, w_m^i)$.
After propagating through the cavity, the outputs corresponding to different tasks are naturally routed to distinct angles, enabling task separation without the need for additional demultiplexers. This architecture simultaneously reduces the system volume and eliminates the insertion loss introduced by demultiplexers, with these advantages becoming increasingly significant as the number of multiplexed channels grows. For each task, the output is collected by a Fourier lens to enhance sensitivity to high-spatial-frequency features~\cite{tancik2020fourier}.

A CMOS camera placed at the focal plane records the optical image via square-law photodetection. The captured images are then downsampled using task-dependent ratios to reduce the dimensionality of the digital NNs. Finally, the digital NNs are trained independently for each task, enabling predictions for all tasks in parallel. Details of the digital NN architectures and training procedures are provided in \textcolor{blue}{Methods}. As shown in \textcolor{blue}{Figure 1b}, our metasurface-based reflection cavity functions as a five-layer all-optical DNN, with only the square-law nonlinearity provided by the camera, similar to the architecture reported in Ref.~\cite{lin2018all}. The key difference is that our metasurface-based DNN contains 115 million fixed, untrained complex-valued parameters. We can therefore combine it with a trainable digital NN to construct a large-scale photonic ELM~\cite{teugin2021scalable,xia2024nonlinear}. By training only the digital NN, the resulting photonic ELM can generalize across various tasks, as demonstrated in this work.

\subsubsection*{Scalable network depth and width boost untrained metasurface DNN performance}

Here, we show that increasing both the diffractive depth (i.e., the number of diffractive layers) and width (i.e., the number of parameters per layer) is critical for improving performance. The benefits of increased depth in DNNs have been discussed in previous studies~\cite{lin2018all,mengu2019analysis} and are further corroborated here by both numerical and experimental results, as shown in \textcolor{blue}{Figures 1c and 1d}. The left panels of \textcolor{blue}{Figure 1c} numerically demonstrate that, as the number of diffractive layers increases, the off-diagonal elements of the transfer matrix become denser, indicating an expansion of the effective parameter space. Physically, each camera pixel receives coherent contributions from a broader range of input modes, generating a richer set of interference-induced quadratic cross-terms. Correspondingly, the output optical field exhibits more randomized speckle patterns with weaker inter-feature correlations, implying a higher effective dimensionality, as shown in the central panels of \textcolor{blue}{Figure 1c}. Together, these effects enhance class separability, as shown in the right panels of \textcolor{blue}{Figure 1c}, thereby improving the classification performance of the trained electronic readout, even though the system nonlinearity remains confined to the final intensity measurement by the camera. Consistent with this interpretation, \textcolor{blue}{Figure 1d} shows that increasing the diffractive depth of the all-optical DNN significantly improves CIFAR-10 classification performance. A detailed explanation of the performance improvement achieved by increasing the number of diffractive layers is provided in \textcolor{blue}{Supplementary Note 2}.

\begin{figure}
\centerline{\includegraphics[trim={0 0 0 0},clip,scale=0.8]{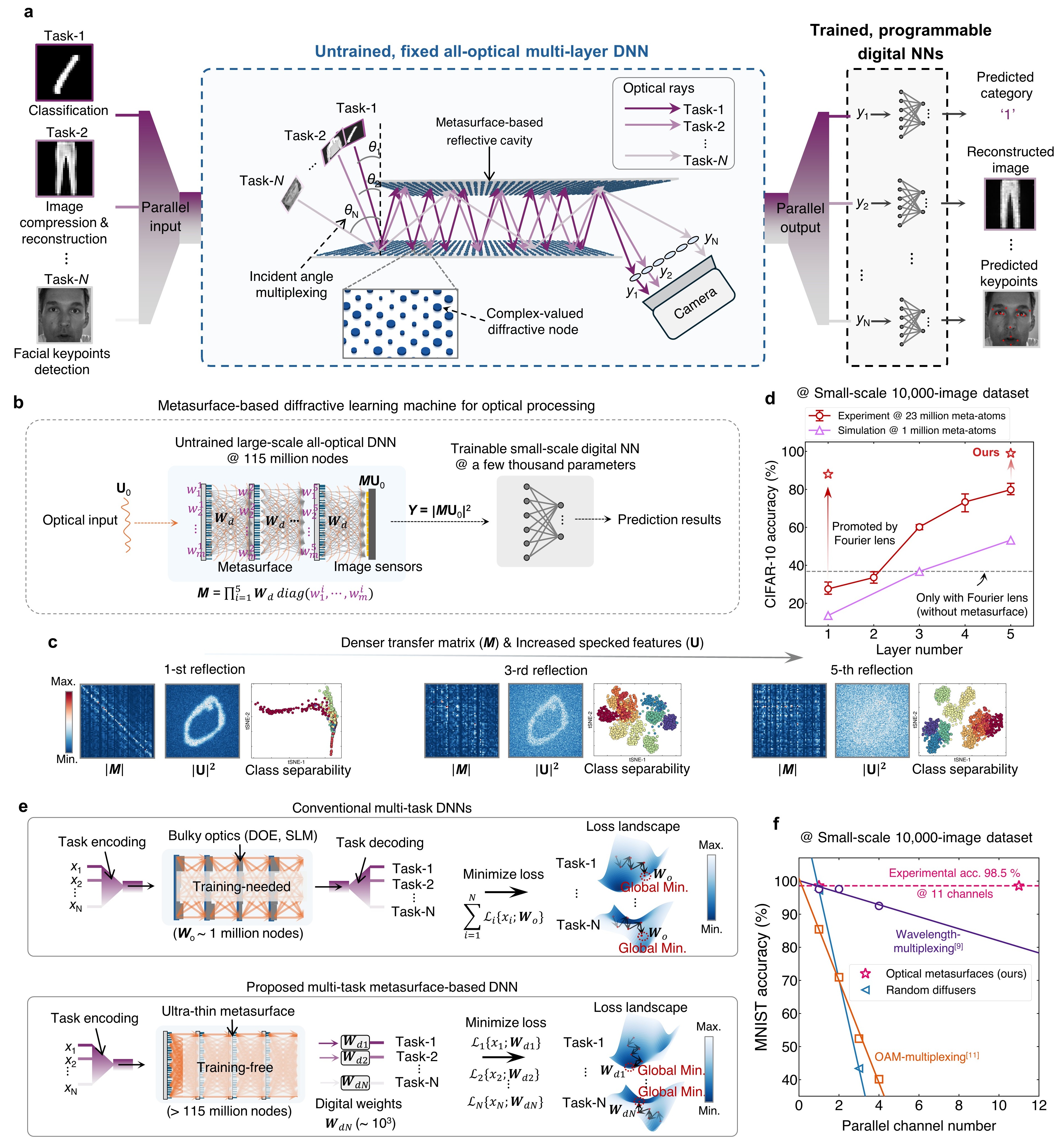}}
\caption{(a) Schematic of the proposed optical-electrical hybrid metasurface-based diffractive learning machine (Meta-DLM) for multi-task deep learning. (b) Architectural illustration of the Meta-DLM consisting of an untrained multi-layer DNN built on the metasurface-based cavity, followed by a trainable digital backend. (c) Transfer matrix and output optical field of the DNN for different numbers of layers. (d) CIFAR-10 accuracy versus the layer number for different DNN configurations. (e) Comparison of the workflow between conventional multi-task DNNs (upper graph) and the proposed multi-task metasurface-based DNN (bottom graph). (f) MNIST accuracy versus the number of parallel channels for different DNN architectures~\cite{duan2023optical, zhang2023advanced}.}
\label{fig1}
\end{figure}

In addition to diffractive depth, the number of parameters plays a key role in improving performance. \textcolor{blue}{Figure 1d} shows that increasing the number of meta-atoms per layer from 1 million to 23 million improves CIFAR-10 classification accuracy across all tested depths. This improvement arises because the 23 million fixed, untrained meta-atoms in each layer project the raw inputs into a higher-dimensional feature space, thereby generating more informative speckle features that enhance classification performance\cite{luo2026highly}. Consequently, the experimental CIFAR-10 accuracy reaches 79.8\% with five layers. Similar trends have been observed in digital ELMs, where increasing the number of hidden neurons improves classification accuracy\cite{tissera2016deep,qing2020deep,soria2011belm,zhou2014stacked}. Moreover, incorporating an optical lens that performs a Fourier transform further increases the experimental five-layer accuracy to 99.3\%. This is because the Fourier lens enables the extraction of high-frequency Fourier features from the fully connected DNN~\cite{tancik2020fourier}. This accuracy is obtained using a small-scale CIFAR-10 dataset containing 10,000 images. We further evaluate our system on the full CIFAR-10 dataset of 60,000 images and achieve a classification accuracy of 95.3\%. Detailed experimental results are presented in Section 2.2.

\subsubsection*{Scalable multi-task processing machanism}

In addition to scalability in network depth and width, our system also supports scalable multi-task inference. In conventionally trained DNNs, multi-task operation typically requires jointly optimizing the diffractive-node parameters across all tasks, as illustrated in the upper panel of \textcolor{blue}{Figure 1e}. This joint training can reduce the overall loss, $\sum_{i=1}^{N}\mathcal{L}_i$, toward a global minimum; however, the loss of each individual task generally cannot reach its own optimum because tasks compete during shared-parameter optimization. This competition becomes more pronounced as the number of tasks increases, ultimately leading to substantial performance degradation. Consistent with this effect, the MNIST accuracy drops dramatically as the number of parallel task channels increases in conventional multi-task DNNs, as shown in \textcolor{blue}{Figure 1f}. 

In contrast, in our system, all tasks share the same untrained DNN with fixed diffractive nodes, while each task is refined by a lightweight, task-specific digital network, as shown in the bottom panel of \textcolor{blue}{Figure 1e}. This decouples optimization across tasks: each loss $\mathcal{L}_i$ can be minimized independently, compensating for the fixed metasurface response and allowing each task to approach its own optimum, thereby maximizing overall system performance. In addition, our system encodes parallel tasks on distinct incident angles. Owing to the resonant nature of meta-atoms, their optical response is strongly angle-dependent~\cite {kamali2017angle,liu2018metasurface}. This angular selectivity naturally spatially separates channels, eliminating the need for additional demultiplexers and thereby reducing insertion loss and system complexity while maintaining scalability. 

As a result, we experimentally achieve a consistent MNIST accuracy of 98.5\% over 11 parallel channels, as shown in \textcolor{blue}{Figure 1f}, representing the largest number of tasks reported to date~\cite{duan2023optical,luo2022metasurface,wang2024matrix,chi2025metasurface,wang2025high,liu2025polarization,liu2025ultra}. Under the same experimental conditions, the number of parallel tasks is theoretically predicted to reach up to 26. The multi-task performance of our system will be fully characterized in Sections 2.3 and 2.4. Additionally, we experimentally demonstrate that, even when using the same incident-angle-multiplexing scheme as our Meta-DLM, conventional random diffusers still exhibit a substantial accuracy drop as the number of parallel channels increases, as shown in \textcolor{blue}{Figure 1f}. This degradation arises from the large angular divergence of random diffusers, which induces pronounced diffraction crosstalk between detection channels and consequently severely degrades multitask performance. Further experimental details are provided in \textcolor{blue}{Supplementary Note 5}.

\subsection{Performance benchmarking with a single task}

We begin by experimentally benchmarking the single-task performance of our Meta-DLM. In the experiment, the single-task performance of Meta-DLM is evaluated on three typical benchmarking datasets (MNIST, Fashion-MNIST, and CIFAR-10). The experimental setup for the single-task characterization is illustrated in \textcolor{blue}{Figure 2a}.  To facilitate easy adjustment of the layer number, we use five individual metasurface chips to create the reflection cavity that acts as a multi-layer all-optical DNN with fixed, untrained diffractive nodes, as shown in \textcolor{blue}{Figure 2b}. In the experiments, the optical image is first generated by an SLM and injected into the metasurface-based cavity. After being processed by the cavity, the optical image is collected by an optical lens and recorded by a CMOS camera. The captured images are then used to train a digital NN with only 7,680 weights, yielding the final classification accuracy. Details of the digital NN architecture and training procedure are provided in \textcolor{blue}{Methods.} More details about the experimental setup are provided in \textcolor{blue}{Methods} and \textcolor{blue}{Supplementary Note 4}.

Here, we experimentally demonstrate that the task performance of our Meta-DLM can be significantly improved by increasing the number of metasurface layers. For comparison, we create a baseline model using only a single-layer metasurface without an optical lens. As shown in \textcolor{blue}{Figures 2d} to \textcolor{blue}{2f}, as the DNN depth increases from one to five layers, the experimental accuracy improves significantly across all tasks. Specifically, the MNIST accuracy improves from 82.9\% to 98.4\%, the Fashion-MNIST accuracy from 83.6\% to 92.6\%, and the CIFAR-10 accuracy from 27.6\% to 79.8\%. The above accuracies are obtained by averaging the results from 10 distinct test datasets. Furthermore, we use the t-SNE method~\cite{vandermaaten08a} to project experimentally captured images into a low-dimensional space for better visualization, as shown in \textcolor{blue}{Figures 2g} and \textcolor{blue}{2h}. We observe from the t-SNE results that data points form more distinct clusters as the layer number increases. This indicates that a deeper metasurface-based DNN can enable a more efficient extraction of category-related features from the raw inputs.

\begin{figure}
\centerline{\includegraphics[trim={0 0 0 0.1cm},clip,scale=0.78]{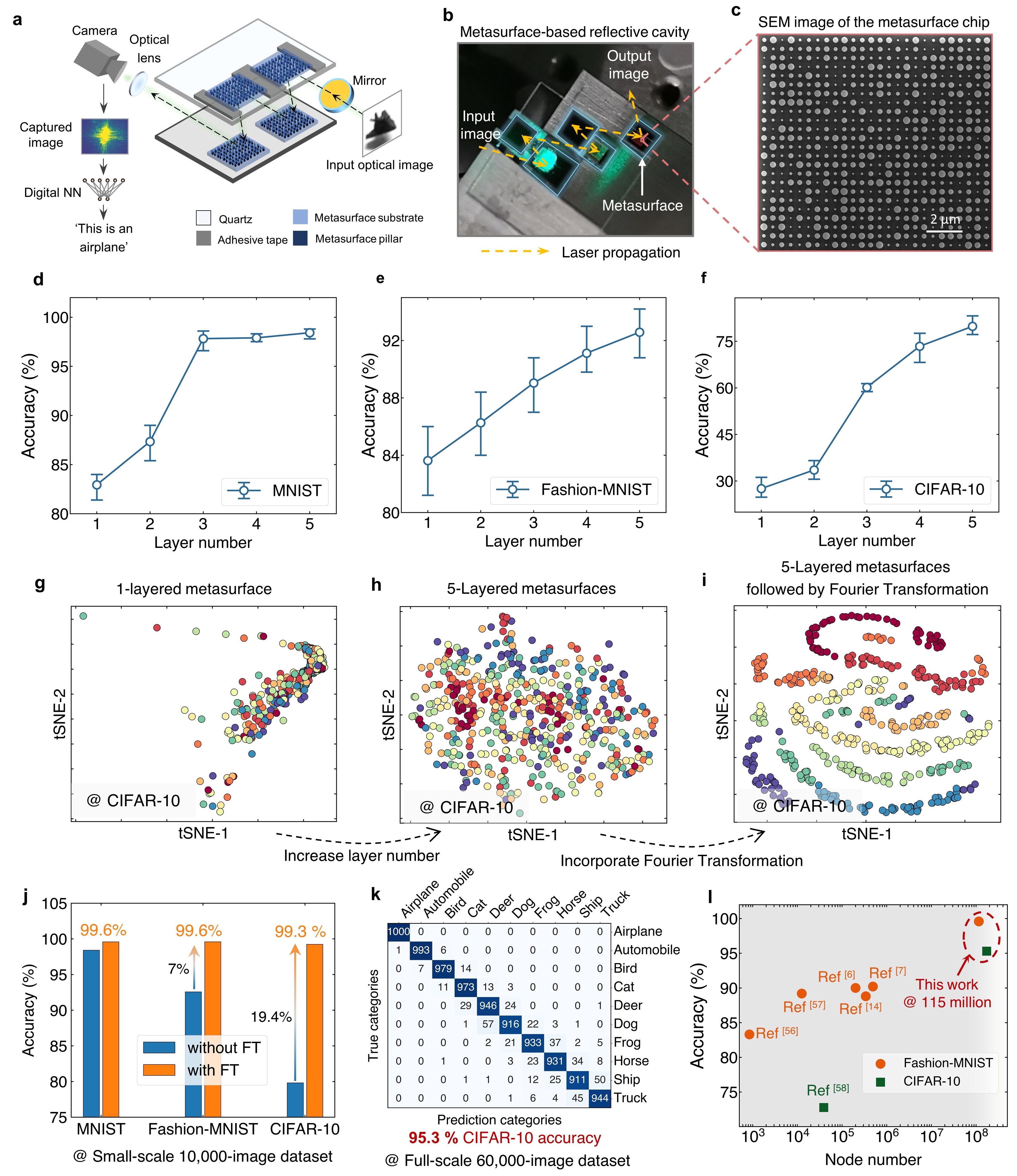}}
\caption{(a) Illustration of the experimental setup for single-task characterization. (b) Photograph of the metasurface-based reflection cavity used in the experiment. (c) Scanning electron microscope (SEM) image of the fabricated metasurface chip. (d)-(f) Experimental accuracy versus the number of metasurface layers for MNIST (d), Fashion-MNIST (e), and CIFAR-10 (f). (g)-(i) t-SNE visualizations of the optical output from Meta-DLM for CIFAR-10, using: (g) a single-layer metasurface, (h) a five-layer metasurface, and (i) a five-layer metasurface followed by the Fourier transform (FT). Colors represent different categories. (j) Accuracy comparison of Meta-DLM with and without FT across the three tasks, each evaluated using a small-scale dataset comprising 10,000 images. (k) Confusion matrix of the CIFAR-10 classification result obtained on the standard full-scale dataset comprising 60,000 images. (l) Accuracy comparison of Meta-DLM with existing all-optical DNNs on three benchmarking tasks~\cite{bernstein2023single,qu2022all,lin2018all,zheng2024multichannel,zhou2021large,wei2024spatially}.}
\label{fig2}
\end{figure}

Additionally, we also demonstrate that the optical lens plays an important role in our system. \textcolor{blue}{Figure 2i} shows that, when an optical lens is added to our system, data points form more distinct clusters in the CIFAR-10 dataset. The accuracy improvement brought by the optical lens in the three tasks is depicted in \textcolor{blue}{Figure 2j}, where the CIFAR-10 accuracy is significantly increased from 79.8\% to 99.3\% on a small-scale 10,000-image dataset. Such significant improvement is because the Fourier transformation performed by the optical lens enhances the ability of the network to learn high-frequency information, which is usually lacking in fully connected networks~\cite{tancik2020fourier}. Furthermore, we experimentally demonstrate that the above superior performance is uniquely enabled by our metasurfaces and cannot be achieved using conventional random diffusers. The random-diffuser-based system experimentally achieves a CIFAR-10 accuracy of only 65.2\% with five layers, substantially lower than the 99.3\% achieved by our Meta-DLM. This performance gap can be attributed to the relatively large scattering-feature sizes of random diffusers, which typically range from a few to tens of micrometers~\cite{mosk2012controlling}, thereby limiting the number of available photonic nodes in each layer. Further experimental details are provided in \textcolor{blue}{Supplementary Note 5}.

To ensure a fair comparison with digital models, which are typically evaluated on the standard full-scale CIFAR-10 dataset comprising 50,000 training images and 10,000 test images, we experimentally evaluate our system using the same dataset. To accommodate the requirements of the optical experiments, all colorful CIFAR-10 images are converted to grayscale. As shown in \textcolor{blue}{Figure 2k}, our system achieves a high classification accuracy of 95.3\% on the standard full-scale 60,000-image CIFAR-10 dataset using only 7,680 digitally trainable parameters. Notably, our Meta-DLM performs comparably with the 96.6\% CIFAR-10 accuracy achieved by a 3.3-million-parameter digital model~\cite{real2019regularized}, which is evaluated under the same grayscale-input and dataset-split conditions. Compared with large-scale digital models evaluated on the standard color CIFAR-10 dataset, our system shows only modest accuracy gaps of 2.6\% and 3.2\% relative to a 26.2-million-parameter ResNet (97.9\%)~\cite{hu2018squeeze} and a 26-million-parameter PyramidNet (98.5\%)~\cite{cubuk2019autoaugment}, respectively. These differences may be partly attributed to the use of grayscale inputs in our optical system, as color information has been reported to improve neural network recognition accuracy on the CIFAR-10 dataset~\cite{zheng2015compact}. \textcolor{blue}{Figure 2l} shows a performance comparison between our system and various state-of-the-art all-optical DNNs. Our system experimentally achieves not only the highest accuracy but also the largest-scale diffractive nodes of up to 115 million, significantly surpassing current all-optical DNNs.

\subsubsection*{Robustness against misalignment errors and fabrication variances}

Here, we experimentally demonstrate that our Meta-DLM can recover quickly from the external perturbation of misalignment errors, which is highly desired in real-time practical applications. To demonstrate this, an axial misalignment of $\sim$ 10 $\mu$m (larger than 10 times the working wavelength) is applied to one of the five metasurface chips. After introducing this error, the MNIST accuracy dramatically drops to 7.8\% from the initial 99\%. We then use the misaligned system to capture a new image dataset, which is used to retrain the compact digital NN for only 40 epochs. As a result, the MNIST accuracy is restored to 98.8\%. With our current experimental setup, the entire recalibration process takes 26.6 s, including optical data acquisition, image downsampling, and retraining of the digital NN. With a high-speed camera and SLM operating at 2,000 frames per second~\cite{pivnenko2021sub,ishii20102000}, the optical data-acquisition time could be further reduced to below 1s in the future. Therefore, digital-backend retraining provides a time-efficient and practically feasible approach for adapting the system to external perturbations while avoiding physical readjustment of the metasurface layers. Detailed experimental results are provided in \textcolor{blue}{Supplementary Note 6.}

In addition to misalignment errors, our Meta-DLM also exhibits strong robustness to fabrication errors. The fabrication process inevitably introduces fabrication errors, which will accumulate as the DNN scale increases, ultimately degrading practical performance. Despite the large-scale parameters of our system, these parameters are randomly assigned from a Gaussian statistical distribution. This enables our system to inherently exhibit strong robustness to fabrication errors, which also randomly occur following a Gaussian distribution~\cite{banerji2020impact}. Simulation results demonstrate that both the phase- and amplitude-modulation coefficients retain approximately Gaussian random distributions when the fabrication errors range from 5\% to 15\% of the mean meta-atom diameter. Consequently, the system maintains highly consistent performance over this error range, with the simulated MNIST classification accuracy varying by only 0.2\%. Simulation details are provided in \textcolor{blue}{Supplementary Notes 1 and 3}.

\subsection{Multi-task parallel processing of different types of tasks}\label{sec2.3}

Encouraged by the remarkable performance in single-task processing, we further evaluate the multi-task performance of our Meta-DLM. To the best of our knowledge, currently reported multi-tasking DNNs only demonstrate simple classification tasks~\cite{duan2023optical,cheng2024photonic,wang2024matrix,luo2022metasurface,zhang2023advanced}; However, practical applications usually involve tasks of diverse types and difficulties. Therefore, we characterize our Meta-DLM's performance in the parallel processing of three distinct task types: image classification, image compression and reconstruction, and object detection. The experimental setup for multi-task characterization is schematized in \textcolor{blue}{Figure 3a}. In the experiments, input optical images ($x_1$,$x_2$, and $x_3$) corresponding to three different tasks are generated by an SLM. These three images are then injected into the metasurface-based reflection cavity at three distinct incident angles ($\theta_1$,$\theta_2$, and $\theta_3$). After being processed by the metasurface reflection cavity, the optical images are spatially separated. The output images are individually collected by three optical lenses and recorded by a CMOS camera. The captured images are then used to train three separate digital NNs, which generate prediction results for the three tasks in parallel. Details of dataset preparation and digital NN training for multi-task processing are provided in \textcolor{blue}{Methods}. Details about the experimental setup are provided in \textcolor{blue}{Supplementary Note 7}.

\begin{figure}
\centerline{\includegraphics[trim={0cm 0cm 0cm 0cm},clip,scale=0.8]{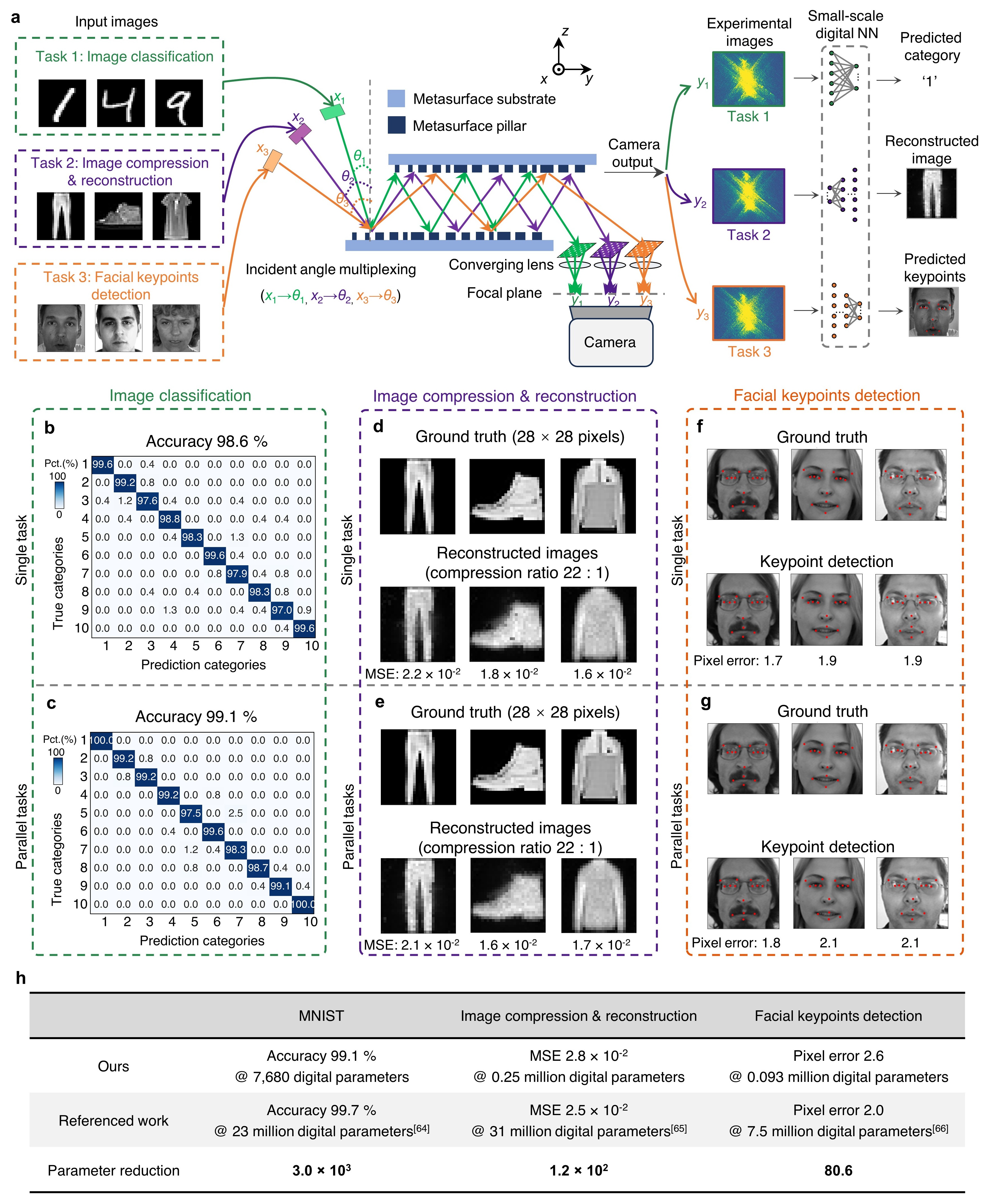}}
\caption{(a) Schematic of our Meta-DLM for the parallel processing of three different types of tasks. (b) and (c) Experimentally obtained MNIST confusion matrices for single-task (b) and multi-task (c) operation. (d) and (e) Original and reconstructed Fashion-MNIST images from single-task (d) and multi-task (e) experiments. (f) and (g) Original facial images and predicted keypoints from single-task (f) and multi-task (g) experiments. (h) Performance comparison between our Meta-DLM and the referenced works~\cite{9066950, Zhu2021ImageRT, Agarwal2017FacialKP}.}
\label{fig3}
\end{figure}
In the multi-task experiment, neighboring channels inevitably cause crosstalk at the detection plane due to light diffraction. To investigate the impact of this crosstalk on performance, we compare multi-task results with single-task benchmarks, which are free from such interference. For image classification, the multi-task MNIST accuracy of 99.2\% is slightly higher than the single-task accuracy of 98.6\%, as shown in \textcolor{blue}{Figures 3b} and \textcolor{blue}{3c}. This improvement occurs because the diffraction-induced crosstalk acts as an additional source of random noise, which can enhance the generalization ability of Meta-DLM. Evidence for this is seen in the significant reduction of the gap between training and testing accuracy, from 4.3\% under single-task conditions to 1.0\% under multi-task conditions. For image compression and reconstruction, the input image with 28 $\times$ 28 pixels is compressed to 6 $\times$ 6 pixels using our Meta-DLM as an optical encoder, with an image compression ratio of $\sim$ 22 : 1. The compressed image is then reconstructed using a 5-layered fully-connected NN as a digital decoder, achieving an average mean square error (MSE) between the reconstructed and original images as low as 2.80 $\times$ 10$^{-2}$ under parallel multiple tasks, as shown in \textcolor{blue}{Figures 3d} and \textcolor{blue}{3e}. This is comparable to the average MSE of 2.72 $\times$ 10$^{-2}$ under a single task. 

For the challenging task of facial keypoint detection, a 9-layer fully-connected NN serves as a digital regression layer. It is trained on experimentally captured 6 $\times$ 6 pixel images to predict 15 coordinates representing facial keypoint positions. As shown in \textcolor{blue}{Figures 3f} and \textcolor{blue}{3g}, the average root mean square error (RMSE) between the ground truth and predicted coordinates is as low as 2.64 under parallel multi-task conditions, showing little difference from the single-task RMSE of 2.62. These results demonstrate that our Meta-DLM can perform multiple tasks of different types in parallel with performance consistent with single-task operation. Remarkably, its multi-task performance rivals that of large-scale digital models dedicated to a single task, as shown in \textcolor{blue}{Figure 3h}. Furthermore, by offloading intensive computation to the optical domain, Meta-DLM reduces the number of required digital parameters by factors ranging from tens to thousands.

\subsection{Highly scalable multi-task processing ability enabled by Meta-DLM}\label{sec2.4}

In this section, we demonstrate that our Meta-DLM can experimentally realize highly scalable multi-task processing, significantly outperforming prior multi-task DNNs. Under the current experimental setup, we experimentally verify that Meta-DLM can theoretically support the parallel processing of up to 26 tasks at most. Furthermore, we experimentally demonstrate the simultaneous processing of 11 tasks, with the performance of each task evaluated independently.

\subsubsection*{Theoretical analysis of the multi-task processing scalability of Meta-DLM}

Here, we theoretically analyze the maximum number of tasks that our Meta-DLM can process in parallel. Under idealized angular-packing assumptions, this limit is determined by how many optical images can be simultaneously injected into the metasurface-based reflection cavity at distinct incident angles. Therefore, the largest number of input channels can be expressed as $N = \theta_{i, max}/\Delta\theta_{min}$, where $\theta_{i, max}$ refers to the maximum range of incident angle ($\theta_i$) and $\Delta\theta_{min}$ refers to the minimum angle difference between two adjacent channels ($\Delta\theta$), as shown in \textcolor{blue}{Figure 4a}. In practice, the achievable number of channels is expected to be lower because of constraints such as the detector field of view and the angular response of the metasurface.  

In the following, we measure the maximum range ($\theta_{i, max}$) and minimum difference ($\Delta\theta_{min}$) of incident angles using the experimental setup shown in \textcolor{blue}{Figure 4b}. More details about the experimental setup can be found in \textcolor{blue}{Supplementary Note 7}. The current experimental setup for proof-of-concept is bulky. Nevertheless, our metasurface-based reflection cavity could potentially be compactly integrated with the CMOS sensor chip and on-board digital processor due to the compact volume of only 216 mm$^3$ in the experiment, as shown in \textcolor{blue}{Figure 4c}. As shown in \textcolor{blue}{Figure 4d}, Meta-DLM maintains a consistent MNIST accuracy of $\sim$ 99\% across all tested angles, along with a consistent facial keypoint error of 3.05 pixels. Therefore, the maximum range of the incident angle ($\theta_{i, max}$) is experimentally measured as 60$^\circ$. As shown in \textcolor{blue}{Figures 4e} and \textcolor{blue}{4f}, consistent performance is also achieved for two parallel tasks across various incident-angle differences, with the minimum angle difference ($\Delta\theta_{min}$) measured as 2.3$^\circ$. Hence, the upper bound on the number of available parallel channels is 26, as estimated using $N = \theta_{i, max}/\Delta\theta_{min}$ = 60$^\circ$/2.3$^\circ$.

\begin{figure}
\centerline{\includegraphics[trim={0cm 0cm 0cm 0cm},clip,scale=0.8]{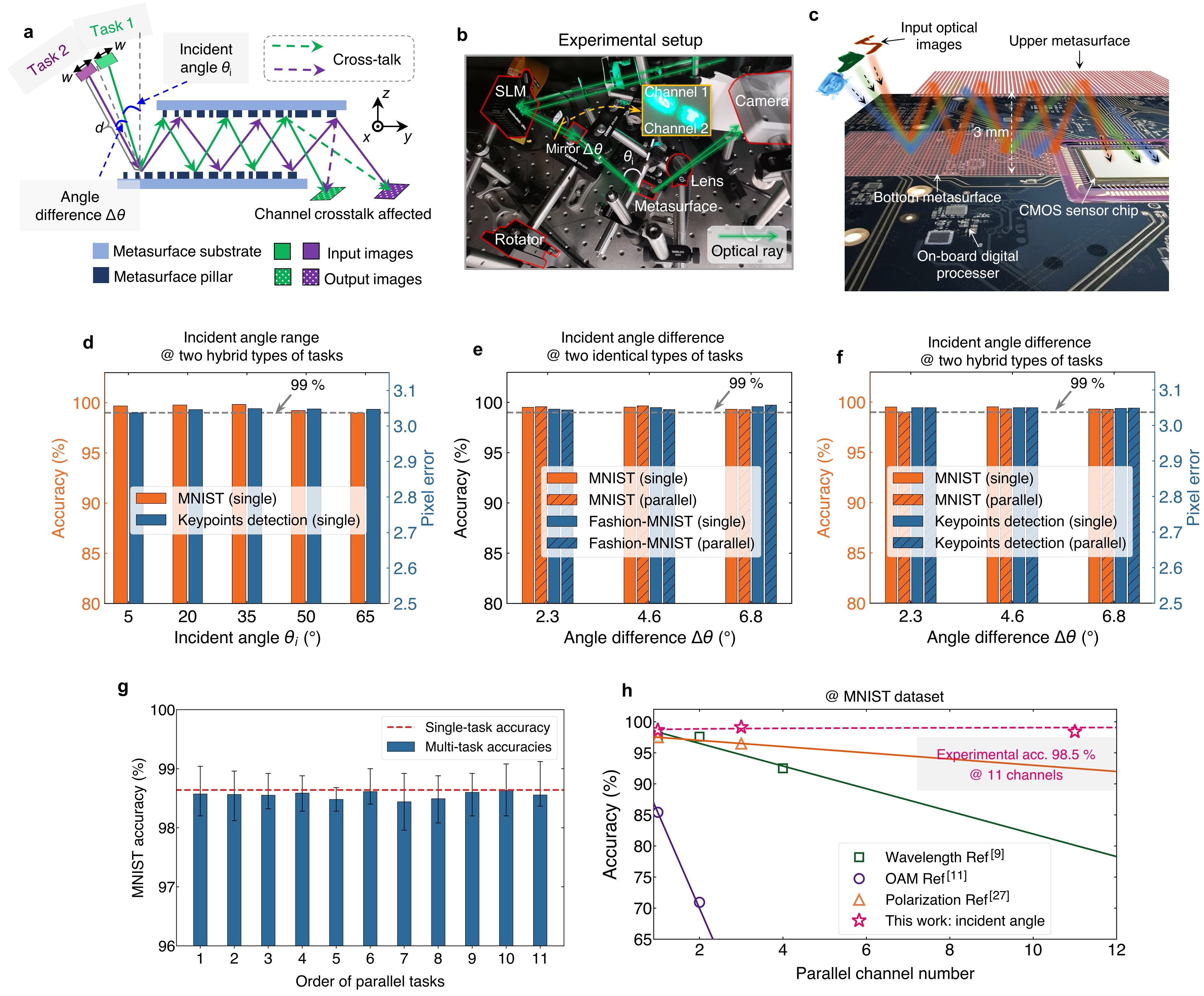}}
\caption{(a) Illustration of Meta-DLM for the parallel processing of two tasks. (b) Photograph of the experimental setup for the parallel processing of two tasks. (c) Schematic of Meta-DLM compactly integrated with the CMOS sensor chip and on-board digital processor. (d) MNIST accuracy and pixel error of Meta-DLM versus incident angle. (e) MNIST and Fashion-MNIST accuracies of Meta-DLM versus incident angle difference. (f) Dependence of MNIST accuracy and facial keypoint error on incident angle difference. (g) The MNIST accuracies of different task channels. (h) MNIST accuracy of Meta-DLM versus existing DNNs as a function of the number of parallel channels~\cite{duan2023optical,zhang2023advanced,wang2024matrix}.}
\label{fig4}
\end{figure}

\subsubsection*{Experimental demonstration of the parallel processing of 11 tasks using Meta-DLM }

In the above experiments, we consider crosstalk from only one neighboring channel. In practical multitask deployments, however, the performance of each task can be affected by crosstalk from all parallel channels due to light diffraction. Therefore, we further evaluate the multitask performance of our system by simultaneously operating 11 task channels and independently evaluating the performance of each channel. In the experiments, each channel is assigned an independently selected sequence of MNIST images and their corresponding ground-truth labels. This ensures that the 11 channels carry independent, time-varying input signals during simultaneous operation. The same MNIST classification task and dataset are used for all channels to ensure a fair comparison of task performance. All 11 MNIST channels are processed simultaneously by the metasurface-based reflection cavity. Because the channel outputs are spatially separated in free space, the camera can independently record the output image from each channel. The recorded output image from each channel is then processed by a separate digital NN, each containing the same 7,680 weights, to obtain the final classification accuracy for that channel. Further experimental details are provided in \textcolor{blue}{Methods}.

As shown in \textcolor{blue}{Figure 4g}, all 11 channels achieve MNIST classification accuracies above 98.5\% during simultaneous operation, comparable to the single-task accuracy of 98.6\%. These channel-resolved results demonstrate that our system can process all 11 tasks in parallel without significant performance degradation in any individual channel. To further quantify the channel-to-channel variation, we calculate the standard deviation of the differences between the classification accuracy of each channel and the mean accuracy across the 11 channels. This value is only 0.06\%, indicating excellent performance uniformity across the channels. We compare the multi-task performance of our system with the state-of-the-art multi-task DNNs, as shown in \textcolor{blue}{Figure 4h}. Our Meta-DLM exhibits a consistent MNIST accuracy of over 98.5\% over various numbers of parallel channels. In contrast, conventional multi-task DNNs experience a significant drop in MNIST accuracy as the number of parallel channels increases. Our system experimentally achieves the largest number of parallel task channels up to 11, while simultaneously reaching the highest MNIST accuracy up to 98.5\%.

\section{Discussion and Conclusion}

We experimentally demonstrate an optical-electrical hybrid metasurface-based diffractive learning machine (Meta-DLM) for angle-multiplexed multi-task processing. This system comprises a fixed, untrained metasurface-based all-optical DNN, coupled with a trainable compact digital layer. The all-optical DNN is created using a double-layer metasurface-based reflection cavity, which allows direct scaling of the DNN’s depth by increasing the number of light reflections, without requiring new metasurface chips for additional layers. Using this design, we experimentally construct a five-layer all-optical DNN, offering 115 million fixed diffractive nodes within a compact volume of 216 mm$^3$. By carefully engineering the optical metasurface, such a large-scale DNN can perform powerful, universal feature extraction via random projections—without requiring any training. As a result, our Meta-DLM can experimentally realize an accuracy of 95.3\% on the standard full-scale CIFAR-10 dataset using only 7,680 digitally trained parameters, significantly outperforming previous DNNs and even rivaling large-scale digital models with millions of parameters. With a novel incident-angle-multiplexing scheme, our Meta-DLM experimentally demonstrates the parallel processing of 11 tasks, each achieving an MNIST accuracy of above 98.5\%, representing the largest-scale multi-task capability among the prior ONNs~\cite{duan2023optical,luo2022metasurface,wang2024matrix,chi2025metasurface,wang2025high,liu2025polarization,liu2025ultra}.

\noindent \textbf{Single-task performance comparison of our system over other untrained ONN systems:} Our system utilizes an optical metasurface-based reflection cavity to realize an untrained DNN with random computing nodes as an optical frontend, coupled with the digital backend to generalize to various tasks. Our system holds one key advantage compared to other untrained systems: the availability of large-scale computing nodes. Conventionally, the untrained optical frontends are usually created by leveraging the programmable randomness of spatial light modulators (SLMs) or the inherent randomness of optical devices, such as optical multi-mode fibers and random scattering media~\cite{ermolaev2025limits,saade2016random,wang2024large,pierangeli2021photonic}. However, the number of their computing nodes is limited compared to ours, thus leading to insufficient learning capability of the NN model. In contrast, optical metasurfaces can provide high-density computing nodes due to the sub-wavelength meta-atoms. This enables our system to achieve 115 million fixed, untrained computing nodes based on a reflection cavity design. Previous demonstrations also use the concept of the reflection cavity to scale the layer number of ONNs~\cite{xia2024nonlinear,yildirim2024nonlinear,dong2025scalable}. However, the computing nodes are still limited by the pixel scale of SLMs, limiting the model scale despite a deeper NN. As a result, our system can experimentally achieve superior task performance, significantly outperforming other ONNs that utilize an untrained optical frontend, as shown in \textcolor{blue}{Table 1}.

%%%% table below
\definecolor{mycolor_0}{RGB}{214,213,193}
\definecolor{mycolor_1}{RGB}{250,250,246}
\definecolor{mycolor_2}{RGB}{236,235,226}

\begin{table}

\caption{The performance comparison of our system with other ONNs based on untrained optical frontends}
% \bigskip
%\begin{center}
%\renewcommand{\arraystretch}{2}
\begin{tabular}{ccccccc}

\hline

\makecell{\textbf{Ref.} }&\makecell{\textbf{Realization}\\ \textbf{of ONNs}} &\makecell{\textbf{Layer} \\ \textbf{number}}& \makecell{\textbf{\# of untrained }\\ \textbf{photonic nodes}} & \makecell{\textbf{Fashion-MNIST} \\ \textbf{accuracy}}& \makecell{\textbf{CIFAR-10} \\ \textbf{accuracy}}&\makecell{\textbf{\# of trained } \\ \textbf{digital parameters}}\\

\hline

\makecell{~\cite{oguz2024programming} \\ ~\cite{ancora2024low} \\ ~\cite{borhani2018learning} \\ ~\cite{ermolaev2025limits} } & \makecell{MMFs} & \makecell{N/A}  & \makecell{A few \\ thousand}& \makecell{ N/A \\N/A \\N/A \\80.0 \% }&  \makecell{N/A }& \makecell{ \textgreater  1 million \\ 70,000 \\ 1,500 \\ 1,440 }\\

\hline

\makecell{~\cite{chen2018imaging} \\ ~\cite{saade2016random} \\ ~\cite{wang2024large}\\ ~\cite{ohana2020kernel} } & \makecell{Random \\ scatterers} & \makecell{Single-layer}  & \makecell{N/A}& \makecell{88.7 \% \\ N/A \\87.0 \% \\88 \%$^1$ }& \makecell{N/A \\ N/A \\ 49.2 \% \\ 79.2 \% }& \makecell{\textgreater 600,000 \\ 100,000 \\ 35,000 \\ \textgreater 60 million}\\

\hline

\makecell{~\cite{pierangeli2021photonic} \\ ~\cite{antonik2019large}  } & \makecell{SLMs} & \makecell{Single-layer}  & \makecell{N/A \\ 16,000}& \makecell{ N/A }& \makecell{ N/A }& \makecell{40,960 \\ 160,000}\\

\hline

\makecell{~\cite{xia2024nonlinear} \\ ~\cite{yildirim2024nonlinear} \\ ~\cite{dong2025scalable}} & \makecell{SLMs + \\ relection layer} & \makecell{\textgreater 5 layers \\ 4 layers \\ 4 layers}  & \makecell{987,840$^*$ \\ 45,000$^{\dag}$ \\ 350,000$^{\dag}$}& \makecell{ 80.0 \% \\ 83.0 \% \\ 65.7 \% }& \makecell{ N/A }& \makecell{10,000 \\ 160 \\ N/A}\\

\hline

\makecell{Ours} & \makecell{Reflection-type \\ metasurfaces} & \makecell{ 5 layers }  & \makecell{115 million}& \makecell{99.6 \%$^2$ }& \makecell{ 95.3 \% }& \makecell{\textless 10,000}\\

\hline

% \multicolumn{8}{l}{Note 1. Operation speed means the prediction or classification speed in photonic RC, and signal processing speed in photonic neural network.}\\
% \multicolumn{8}{l}{Note 2. Energy for generating and detecting optical signals is not included here.}\\
% \multicolumn{8}{l}{Note 3. Metaparameters are the parameters that need to be adjusted in the photonic reservoir layer.}\\
% \multicolumn{8}{l}{Note 4. N/A represents no available data.}
\label{table1}
\end{tabular}
\noindent \small{MMFs and SLMs denote multi-mode fibers and spatial light modulators, respectively}\\
\small{$^1$The accuracy obtained using 100,000 digital parameters}\\
\small{$^2$The accuracy obtained using a small-scale 10,000-image dataset}\\
\small{$^*$The binary-valued photonic nodes enabled with a digital micromirror device (DMD)
}\\
\small{$^{\dag}$The parameters have been end-to-end trained}
% \footnotetext[]{N/A represents no available data.}
\end{table}

\noindent \textbf{Multi-task performance comparison of our system over transfer learning methods:} By just re-training the digital backend for each task, our metasurface-based DNN can be adopted for multiple different tasks, without deploying additional new DNNs. Our approach shares similarities with the transfer learning methods for multi-task learning. In a transfer learning multi-task NN, a frontend network is pre-trained on a large dataset to learn general features, and a smaller backend network is subsequently fine-tuned for specific tasks. Recently, this transfer learning method has been used to create a metasurface-based ONN for multi-task processing~\cite{choi2025transferable}. However, such an end-to-end trained NN extracts features that still highly depend on the chosen dataset, leading to weaker generalization as task diversity grows. In contrast, our approach uses a training-free, 115-million-parameter DNN, which randomly projects the inputs into a universal feature dimension, thus exhibiting better generalization to different datasets. As a result, our system can be experimentally adopted to three different types of tasks. By comparison, the transfer learning-based ONN is limited to the parallel processing of classification tasks and achieves a CIFAR-10 accuracy below 75\% (ours 95.3\%).

\noindent \textbf{Further compression of the system volume:} Another advantage of our system is the availability of the incident-angle-multiplexing scheme, which eliminates the requirement for additional demultiplexers to distinguish different channels. This facilitates the compact integration of the optical frontend with the subsequent digital backend processor. In the current implementation, further miniaturization of the Meta-DLM is constrained by the bulky optical components—the converging lens and reflection mirror—required to generate the various incident angles. As the number of parallel tasks increases, more reflection mirrors and lenses would be required to generate the necessary incident angles. This demand for additional components would exacerbate the system's bulk, thereby limiting the practical applicability of the Meta-DLM. Prior research has demonstrated that engineered phase gradients on a metasurface can precisely control the deflection direction of incident light~\cite{juliano2022metasurface,gao2019highly}. Therefore, by replacing those bulky optics with optical metasurfaces, the system volume of our Meta-DLM can be further compressed in the future.
%%%%%%%%%%%%%%%% REFERENCES %%%%%%%%%%%%%%%
\section{Methods}

\textbf{\emph{The design flow and fabrication of the metasurface}}

1. Design flow: The phase-modulation coefficient of each meta-atom is randomly sampled from a Gaussian distribution with a mean of 0 and a standard deviation of 0.4$\pi$. Based on the complex-valued responses of meta-atoms with different diameters obtained from finite-difference time-domain (FDTD) simulations, the phase- and amplitude-modulation coefficients are interdependent. Once the phase coefficients are selected, the corresponding amplitude coefficients are uniquely determined by the simulated meta-atom responses rather than being designed independently. Hence, the diameter of each meta-atom is determined by the selected phase-modulation coefficient, thereby defining the final layout of the metasurface chip. Further details of the design procedure are provided in \textcolor{blue}{Supplementary Note 1}. 

2. Fabrication: The fabrication of the metasurface starts with a silicon-on-insulator (SOI) wafer, with a 220-nm-height silicon device layer, a 2-$\mu$m-height SiO$_2$ buried layer, and a 725-$\mu$m-thick Si substrate. A layer of electron-resist is spin-coated on the top of the SOI wafer. Electron beam photolithography is then used to define the pattern of meta-atoms. The wafer is then flushed with solutions to ensure that the electron-resist in the region of meta-atoms is preserved while the electron-resist in the other region is removed. Reactive ion etching is used to remove the silicon device layer until the buried layer is exposed to the air. Due to the existence of electron-resist, the silicon device layer with the designed pattern is formed.

\noindent\textbf{\emph{Experimental setup for Meta-DLM}}

For the single-tasking processing, the experimental setup starts with a 532-nm laser diode that generates a collimated laser beam with a diameter of $\sim$ 5 mm. The laser beam illuminates the spatial light modulator (SLM, Meadowlark Optics E19x12-400-800-HDM8) with a pixel size of 8 $\mu$m, which is sandwiched between two linear polarizers with the same polarization directions. By loading a phase pattern with a pixel scale to the SLM, the output laser beam can carry the information of the input image. The phase pattern has a pixel scale of 300 $\times$ 300, so the size of the input optical image of 2.4 $\times$ 2.4 mm$^2$. An aperture is used to block stray light outside of the optical image, as the full area of the laser beam is larger than the optical image. The propagation direction of the optical image is then changed by a reflection mirror. This allows the optical image to be projected to the metasurface chip with an incident angle of $\sim$ 30$^\circ$. The optical image is then reflected by the other 4 metasurface chips in sequence. The spacing distance between the bottom and upper metasurface chips is $\sim$ 5 mm. The reflected image is focused by a converging lens with a focal length of 150 mm. A camera (IRVI Contour IR Digital CMOS Camera) with a pixel resolution of 720 $\times$ 960 and 8-bit depth is placed at the focal plane of the converging lens to capture the optical image. More details about the experimental setup can be found in \textcolor{blue}{Supplementary Note 4}.

For the multiple-tasking parallel processing, the experimental setup is the same as that for the single-task processing, except for the additional components for creating different incident angles. In the experiment, we used two methods to create different incident angles for different tasks. The first way to create the incident angle is using a converging lens. This approach is used in the parallel processing of three parallel tasks, as well as the parallel processing of 11 MNIST tasks. The converging lens can generate a deflection angle dependent on the position of the input light. The input images deflected by the converging lens will propagate with different angles and eventually overlap at the metasurface chip, realizing the angle-multiplexing parallel processing of input optical images. In the experiment, the focal length and diameter of the converging lens are 50 mm and 25.4 mm, respectively. The second way to create the incident angle is by using the reflection mirror mounted on a rotator when only two tasks are processed in parallel. This rotator can control the direction of the reflection mirror and thus change the propagation direction of the input image precisely. Therefore, this approach is used in the study of how the incident angle difference influences the performance of Meta-DLM. More details about the experimental setup can be found in \textcolor{blue}{Supplementary Note 7}.

\noindent\textbf{\emph{Dataset preparation and digital neural network training}}

1. Dataset preparation: For three benchmarking tasks (MNIST, Fashion-MNIST, and CIFAR-10), a total of 10,000 images are randomly selected from the standard full-scale dataset to create a new dataset. Here, 7,500 images are used for model training, while the remaining 2,500 images form an independent evaluation set. The colorful CIFAR images are converted into grayscale. For the task of image compression and reconstruction, 1,500 images and 500 images are randomly selected from the whole Fashion-MNIST dataset to construct the training dataset and testing dataset, respectively. For the facial keypoint detection task, the dataset consists of 2,000 images, of which 1,800 are used for training and 200 for testing. The dataset for facial keypoints detection can be found at https://www.kaggle.com/competitions/facial-keypoints-detection. For all tasks, the dataset image is upsampled to a pixel scale of 300 $\times$ 300 using the nearest area interpolation. The upsampled image is normalized using the maximum pixel value. The normalized image is used to generate a phase pattern to be loaded to the SLM.

2. Digital neural network (NN) training: For the three benchmarking classification tasks, the experimentally captured image is downsampled from 720 $\times$ 960 to 24 $\times$ 32 pixels using pixel binning. The downsampled image is then flattened into a vector of 1 $\times$ 768 and fed into a fully connected digital NN. This NN consists of 768 input nodes and 10 output nodes, with a nonlinear function of $Sigmoid$. During training, categorical cross-entropy is used as the loss function. For the image compression and reconstruction task and the facial keypoint detection task, the experimentally captured image is downsampled to 6 $\times$ 6 pixels using pixel binning. The downsampled image is then flattened into a vector of 1 $\times$ 36 and fed into a multilayer fully connected digital NN. Each layer of this digital NN uses a nonlinear function of $ReLu$. The digital NN consists of five layers for image compression and reconstruction and nine layers for facial keypoint detection. The mean squared error is used as the loss function. For all tasks, the learning rate and batch size are 1 $\times$ 10$^{-3}$ and 32, respectively, and stochastic gradient descent is used as the optimizer. All codes for digital NN training are implemented in Python 3.9 and run on an NVIDIA RTX 3090.

\clearpage % Clear all remaining figures and tables then start a new page

% The list of references goes after the main text and before the acknowledgements
% When preparing an initial submission, we recommend you use BibTeX, like this:
%
\bibliography{main} % for a file named science_template.bib

@Misc{methods,
  note = {Materials and methods are available as supplementary material},
}

@article{luo2026highly,
  title={Highly scalable machine vision enabled with meta-optics-based ultra-wide neural network},
  author={Luo, Mingcheng and Jiang, Meirui and Shastri, Bhavin J and Zhou, Nansen and Guo, Wenfei and Xiong, Jianmin and Wang, Dongliang and Zhou, Renjie and Shu, Chester and Dou, Qi and others},
  journal={eLight},
  volume={6},
  number={1},
  pages={10},
  year={2026},
  publisher={Springer}
}

@article{peng2026optical,
  title={Optical metasurfaces for general vision processing on the edge},
  author={Peng, Jiayong and Luo, Mingcheng and Han, Yuxi and Wu, Siying and Li, Hongsheng and Shastri, Bhavin J and Shu, Chester and Dou, Qi and Chai, Yang and Huang, Chaoran},
  journal={Nature},
  volume={654},
  number={8120},
  pages={917--925},
  year={2026},
  publisher={Nature Publishing Group}
}

@article{antonik2019large,
  title={Large-scale spatiotemporal photonic reservoir computer for image classification},
  author={Antonik, Piotr and Marsal, Nicolas and Rontani, Damien},
  journal={IEEE Journal of Selected Topics in Quantum Electronics},
  volume={26},
  number={1},
  pages={1--12},
  year={2019},
  publisher={IEEE}
}

@article{choi2025transferable,
  title={Transferable polychromatic optical encoder for neural networks},
  author={Choi, Minho and Xiang, Jinlin and Wirth-Singh, Anna and Baek, Seung-Hwan and Shlizerman, Eli and Majumdar, Arka},
  journal={Nature Communications},
  volume={16},
  number={1},
  pages={5623},
  year={2025},
  publisher={Nature Publishing Group UK London}
}

@article{dong2025scalable,
  title={Scalable multilayer diffractive neural network with all-optical nonlinear activation},
  author={Dong, Yiying and Zhang, Bohan and Liang, Ruiqi and Jia, Wenhe and Chen, Kunpeng and Zou, Junye and Hu, Futai and Liu, Sheng and Li, Xiaokai and Yang, Yuanmu},
  journal={arXiv preprint arXiv:2504.13518},
  year={2025}
}

@article{yildirim2024nonlinear,
  title={Nonlinear processing with linear optics},
  author={Yildirim, Mustafa and Dinc, Niyazi Ulas and Oguz, Ilker and Psaltis, Demetri and Moser, Christophe},
  journal={Nature Photonics},
  volume={18},
  number={10},
  pages={1076--1082},
  year={2024},
  publisher={Nature Publishing Group UK London}
}

@article{xia2024nonlinear,
  title={Nonlinear optical encoding enabled by recurrent linear scattering},
  author={Xia, Fei and Kim, Kyungduk and Eliezer, Yaniv and Han, SeungYun and Shaughnessy, Liam and Gigan, Sylvain and Cao, Hui},
  journal={Nature Photonics},
  volume={18},
  number={10},
  pages={1067--1075},
  year={2024},
  publisher={Nature Publishing Group UK London}
}

@article{oguz2024programming,
  title={Programming nonlinear propagation for efficient optical learning machines},
  author={Oguz, Ilker and Hsieh, Jih-Liang and Dinc, Niyazi Ulas and Te{\u{g}}in, U{\u{g}}ur and Yildirim, Mustafa and Gigli, Carlo and Moser, Christophe and Psaltis, Demetri},
  journal={Advanced Photonics},
  volume={6},
  number={1},
  pages={016002--016002},
  year={2024},
  publisher={Society of Photo-Optical Instrumentation Engineers}
}

@article{ancora2024low,
  title={Low-power multimode-fiber projector outperforms shallow-neural-network classifiers},
  author={Ancora, Daniele and Negri, Matteo and Gianfrate, Antonio and Trypogeorgos, Dimitris and Dominici, Lorenzo and Sanvitto, Daniele and Ricci-Tersenghi, Federico and Leuzzi, Luca},
  journal={Physical Review Applied},
  volume={21},
  number={6},
  pages={064027},
  year={2024},
  publisher={APS}
}

@article{borhani2018learning,
  title={Learning to see through multimode fibers},
  author={Borhani, Navid and Kakkava, Eirini and Moser, Christophe and Psaltis, Demetri},
  journal={Optica},
  volume={5},
  number={8},
  pages={960--966},
  year={2018},
  publisher={Optical Society of America}
}

@article{ermolaev2025limits,
  title={Limits of nonlinear and dispersive fiber propagation for an optical fiber-based extreme learning machine},
  author={Ermolaev, Andrei V and Hary, Mathilde and Leybov, Lev and Ryczkowski, Piotr and Skalli, Anas and Brunner, Daniel and Genty, Go{\"e}ry and Dudley, John M},
  journal={Optics Letters},
  volume={50},
  number={13},
  pages={4166--4169},
  year={2025},
  publisher={Optica Publishing Group}
}

@inproceedings{ohana2020kernel,
  title={Kernel computations from large-scale random features obtained by optical processing units},
  author={Ohana, Ruben and Wacker, Jonas and Dong, Jonathan and Marmin, S{\'e}bastien and Krzakala, Florent and Filippone, Maurizio and Daudet, Laurent},
  booktitle={ICASSP 2020-2020 IEEE international conference on acoustics, speech and signal processing (ICASSP)},
  pages={9294--9298},
  year={2020},
  organization={IEEE}
}

@article{wang2024large,
  title={Large-scale photonic computing with nonlinear disordered media},
  author={Wang, Hao and Hu, Jianqi and Morandi, Andrea and Nardi, Alfonso and Xia, Fei and Li, Xuanchen and Savo, Romolo and Liu, Qiang and Grange, Rachel and Gigan, Sylvain},
  journal={Nature Computational Science},
  volume={4},
  number={6},
  pages={429--439},
  year={2024},
  publisher={Nature Publishing Group US New York}
}

@inproceedings{saade2016random,
  title={Random projections through multiple optical scattering: Approximating kernels at the speed of light},
  author={Saade, Alaa and Caltagirone, Francesco and Carron, Igor and Daudet, Laurent and Dr{\'e}meau, Ang{\'e}lique and Gigan, Sylvain and Krzakala, Florent},
  booktitle={2016 IEEE International Conference on Acoustics, Speech and Signal Processing (ICASSP)},
  pages={6215--6219},
  year={2016},
  organization={IEEE}
}

@article{chen2018imaging,
  title={Imaging through scattering media using speckle pattern classification based support vector regression},
  author={Chen, Hui and Gao, Yesheng and Liu, Xingzhao and Zhou, Zhixin},
  journal={Optics express},
  volume={26},
  number={20},
  pages={26663--26678},
  year={2018},
  publisher={Optical Society of America}
}

@article{banerji2020impact,
  title={Impact of fabrication errors and refractive index on multilevel diffractive lens performance},
  author={Banerji, Sourangsu and Cooke, Jacqueline and Sensale-Rodriguez, Berardi},
  journal={Scientific reports},
  volume={10},
  number={1},
  pages={14608},
  year={2020},
  publisher={Nature Publishing Group UK London}
}

@article{zhou2014stacked,
  title={Stacked extreme learning machines},
  author={Zhou, Hongming and Huang, Guang-Bin and Lin, Zhiping and Wang, Han and Soh, Yeng Chai},
  journal={IEEE transactions on cybernetics},
  volume={45},
  number={9},
  pages={2013--2025},
  year={2014},
  publisher={IEEE}
}

@article{soria2011belm,
  title={BELM: Bayesian extreme learning machine},
  author={Soria-Olivas, Emilio and Gomez-Sanchis, Juan and Martin, Jos{\'e} D and Vila-Frances, Joan and Martinez, Marcelino and Magdalena, Jos{\'e} R and Serrano, Antonio J},
  journal={IEEE transactions on neural networks},
  volume={22},
  number={3},
  pages={505--509},
  year={2011},
  publisher={IEEE}
}

@article{qing2020deep,
  title={Deep and wide feature based extreme learning machine for image classification},
  author={Qing, Yuanyuan and Zeng, Yijie and Li, Yue and Huang, Guang-Bin},
  journal={Neurocomputing},
  volume={412},
  pages={426--436},
  year={2020},
  publisher={Elsevier}
}

@article{tissera2016deep,
  title={Deep extreme learning machines: supervised autoencoding architecture for classification},
  author={Tissera, Migel D and McDonnell, Mark D},
  journal={Neurocomputing},
  volume={174},
  pages={42--49},
  year={2016},
  publisher={Elsevier}
}

@article{ma2020artificial,
  title={Artificial intelligence applications in the development of autonomous vehicles: A survey},
  author={Ma, Yifang and Wang, Zhenyu and Yang, Hong and Yang, Lin},
  journal={IEEE/CAA Journal of Automatica Sinica},
  volume={7},
  number={2},
  pages={315--329},
  year={2020},
  publisher={IEEE}
}

@article{lee2021application,
  title={Application of artificial intelligence-based technologies in the healthcare industry: Opportunities and challenges},
  author={Lee, DonHee and Yoon, Seong No},
  journal={International journal of environmental research and public health},
  volume={18},
  number={1},
  pages={271},
  year={2021},
  publisher={MDPI}
}

@article{jacobsen2004research,
  title={Research robots for applications in artificial intelligence, teleoperation and entertainment},
  author={Jacobsen, Stephen C and Olivier, M and Smith, FM and Knutti, David F and Johnson, R Todd and Colvin, GE and Scroggin, WB},
  journal={The International Journal of Robotics Research},
  volume={23},
  number={4-5},
  pages={319--330},
  year={2004},
  publisher={SAGE Publications}
}

@article{min2023recent,
  title={Recent advances in natural language processing via large pre-trained language models: A survey},
  author={Min, Bonan and Ross, Hayley and Sulem, Elior and Veyseh, Amir Pouran Ben and Nguyen, Thien Huu and Sainz, Oscar and Agirre, Eneko and Heintz, Ilana and Roth, Dan},
  journal={ACM Computing Surveys},
  volume={56},
  number={2},
  pages={1--40},
  year={2023},
  publisher={ACM New York, NY}
}

@inproceedings{hacker2023regulating,
  title={Regulating ChatGPT and other large generative AI models},
  author={Hacker, Philipp and Engel, Andreas and Mauer, Marco},
  booktitle={Proceedings of the 2023 ACM conference on fairness, accountability, and transparency},
  pages={1112--1123},
  year={2023}
}

@article{mengu2019analysis,
  title={Analysis of diffractive optical neural networks and their integration with electronic neural networks},
  author={Mengu, Deniz and Luo, Yi and Rivenson, Yair and Ozcan, Aydogan},
  journal={IEEE Journal of Selected Topics in Quantum Electronics},
  volume={26},
  number={1},
  pages={1--14},
  year={2019},
  publisher={IEEE}
}

@article{bernstein2023single,
  title={Single-shot optical neural network},
  author={Bernstein, Liane and Sludds, Alexander and Panuski, Christopher and Trajtenberg-Mills, Sivan and Hamerly, Ryan and Englund, Dirk},
  journal={Science Advances},
  volume={9},
  number={25},
  pages={eadg7904},
  year={2023},
  publisher={American Association for the Advancement of Science}
}

@article{lin2018all,
  title={All-optical machine learning using diffractive deep neural networks},
  author={Lin, Xing and Rivenson, Yair and Yardimci, Nezih T and Veli, Muhammed and Luo, Yi and Jarrahi, Mona and Ozcan, Aydogan},
  journal={Science},
  volume={361},
  number={6406},
  pages={1004--1008},
  year={2018},
  publisher={American Association for the Advancement of Science}
}

@article{wei2024spatially,
  title={Spatially varying nanophotonic neural networks},
  author={Wei, Kaixuan and Li, Xiao and Froech, Johannes and Chakravarthula, Praneeth and Whitehead, James and Tseng, Ethan and Majumdar, Arka and Heide, Felix},
  journal={Science Advances},
  volume={10},
  number={45},
  pages={eadp0391},
  year={2024},
  publisher={American Association for the Advancement of Science}
}

@article{qu2022all,
  title={All-dielectric metasurface empowered optical-electronic hybrid neural networks},
  author={Qu, Geyang and Cai, Guiyi and Sha, Xinbo and Chen, Qinmiao and Cheng, Jiaping and Zhang, Yao and Han, Jiecai and Song, Qinghai and Xiao, Shumin},
  journal={Laser \& Photonics Reviews},
  volume={16},
  number={10},
  pages={2100732},
  year={2022},
  publisher={Wiley Online Library}
}

@article{zhou2021large,
  title={Large-scale neuromorphic optoelectronic computing with a reconfigurable diffractive processing unit},
  author={Zhou, Tiankuang and Lin, Xing and Wu, Jiamin and Chen, Yitong and Xie, Hao and Li, Yipeng and Fan, Jingtao and Wu, Huaqiang and Fang, Lu and Dai, Qionghai},
  journal={Nature Photonics},
  volume={15},
  number={5},
  pages={367--373},
  year={2021},
  publisher={Nature Publishing Group UK London}
}

@article{luo2025large,
  title={Large-scale artificial intelligence with 41 million nanophotonic neurons on a metasurface},
  author={Luo, Mingcheng and Jiang, Meirui and Shastri, Bhavin J and Zhou, Nansen and Guo, Wenfei and Xiong, Jianmin and Wang, Dongliang and Zhou, Renjie and Shu, Chester and Dou, Qi and others},
  journal={arXiv preprint arXiv:2504.20416},
  year={2025}
}

@article{wang2024matrix,
  title={Matrix diffractive deep neural networks merging polarization into meta-devices},
  author={Wang, Yuzhong and Yu, Axiang and Cheng, Yayun and Qi, Jiaran},
  journal={Laser \& Photonics Reviews},
  volume={18},
  number={2},
  pages={2300903},
  year={2024},
  publisher={Wiley Online Library}
}

@article{liu2025ultra,
  title={Ultra-compact multi-task processor based on in-memory optical computing},
  author={Liu, Wencan and Huang, Yuyao and Sun, Run and Fu, Tingzhao and Yang, Sigang and Chen, Hongwei},
  journal={Light: Science \& Applications},
  volume={14},
  number={1},
  pages={134},
  year={2025},
  publisher={Nature Publishing Group UK London}
}

@article{liu2025polarization,
  title={Polarization-multiplexed diffractive neural networks for multi-task classification based on liquid crystals},
  author={Liu, Mengqin and Ye, Xianglin and Zhou, Yingjie and Tang, Dongliang and Fan, Fan},
  journal={Optics Letters},
  volume={50},
  number={17},
  pages={5446--5449},
  year={2025},
  publisher={Optica Publishing Group}
}

@article{wang2025high,
  title={High-capacity directional information processor using all-optical multilayered neural networks},
  author={Wang, Guannan and Zang, Xiaofei and Zhang, Teng and Tan, Zhiyu and Guo, Ziqing and Zhu, Yiming and Chen, Xianzhong and Zhuang, Songlin},
  journal={Science Advances},
  volume={11},
  number={47},
  pages={eadu0904},
  year={2025},
  publisher={American Association for the Advancement of Science}
}

@article{chi2025metasurface,
  title={Metasurface enabled multi-target and multi-wavelength diffraction neural networks},
  author={Chi, Haoxiang and Zang, Xiaofei and Zhang, Teng and Wang, Guannan and Fan, Zhiyuan and Zhu, Yiming and Chen, Xianzhong and Zhuang, Songlin},
  journal={Laser \& Photonics Reviews},
  volume={19},
  number={1},
  pages={2401178},
  year={2025},
  publisher={Wiley Online Library}
}

@article{luo2022metasurface,
  title={Metasurface-enabled on-chip multiplexed diffractive neural networks in the visible},
  author={Luo, Xuhao and Hu, Yueqiang and Ou, Xiangnian and Li, Xin and Lai, Jiajie and Liu, Na and Cheng, Xinbin and Pan, Anlian and Duan, Huigao},
  journal={Light: Science \& Applications},
  volume={11},
  number={1},
  pages={158},
  year={2022},
  publisher={Nature Publishing Group UK London}
}

@article{zheng2022meta,
  title={Meta-optic accelerators for object classifiers},
  author={Zheng, Hanyu and Liu, Quan and Zhou, You and Kravchenko, Ivan I and Huo, Yuankai and Valentine, Jason},
  journal={Science Advances},
  volume={8},
  number={30},
  pages={eabo6410},
  year={2022},
  publisher={American Association for the Advancement of Science}
}

@article{zheng2024multichannel,
  title={Multichannel meta-imagers for accelerating machine vision},
  author={Zheng, Hanyu and Liu, Quan and Kravchenko, Ivan I and Zhang, Xiaomeng and Huo, Yuankai and Valentine, Jason G},
  journal={Nature nanotechnology},
  volume={19},
  number={4},
  pages={471--478},
  year={2024},
  publisher={Nature Publishing Group UK London}
}

@article{li2019intelligent,
  title={Intelligent metasurface imager and recognizer},
  author={Li, Lianlin and Shuang, Ya and Ma, Qian and Li, Haoyang and Zhao, Hanting and Wei, Menglin and Liu, Che and Hao, Chenglong and Qiu, Cheng-Wei and Cui, Tie Jun},
  journal={Light: science \& applications},
  volume={8},
  number={1},
  pages={97},
  year={2019},
  publisher={Nature Publishing Group UK London}
}

@article{duan2023optical,
  title={Optical multi-task learning using multi-wavelength diffractive deep neural networks},
  author={Duan, Zhengyang and Chen, Hang and Lin, Xing},
  journal={Nanophotonics},
  volume={12},
  number={5},
  pages={893--903},
  year={2023},
  publisher={De Gruyter}
}

@article{cheng2024photonic,
  title={Photonic neuromorphic architecture for tens-of-task lifelong learning},
  author={Cheng, Yuan and Zhang, Jianing and Zhou, Tiankuang and Wang, Yuyan and Xu, Zhihao and Yuan, Xiaoyun and Fang, Lu},
  journal={Light: Science \& Applications},
  volume={13},
  number={1},
  pages={56},
  year={2024},
  publisher={Nature Publishing Group UK London}
}

@article{zhang2023advanced,
  title={Advanced all-optical classification using orbital-angular-momentum-encoded diffractive networks},
  author={Zhang, Kuo and Liao, Kun and Cheng, Haohang and Feng, Shuai and Hu, Xiaoyong},
  journal={Advanced Photonics Nexus},
  volume={2},
  number={6},
  pages={066006--066006},
  year={2023},
  publisher={Society of Photo-Optical Instrumentation Engineers}
}

@article{luo2019design,
  title={Design of task-specific optical systems using broadband diffractive neural networks},
  author={Luo, Yi and Mengu, Deniz and Yardimci, Nezih T and Rivenson, Yair and Veli, Muhammed and Jarrahi, Mona and Ozcan, Aydogan},
  journal={Light: Science \& Applications},
  volume={8},
  number={1},
  pages={112},
  year={2019},
  publisher={Nature Publishing Group UK London}
}

@article{chen2016review,
  title={A review of metasurfaces: physics and applications},
  author={Chen, Hou-Tong and Taylor, Antoinette J and Yu, Nanfang},
  journal={Reports on progress in physics},
  volume={79},
  number={7},
  pages={076401},
  year={2016},
  publisher={IOP Publishing}
}

@article{overvig2019dielectric,
  title={Dielectric metasurfaces for complete and independent control of the optical amplitude and phase},
  author={Overvig, Adam C and Shrestha, Sajan and Malek, Stephanie C and Lu, Ming and Stein, Aaron and Zheng, Changxi and Yu, Nanfang},
  journal={Light: Science \& Applications},
  volume={8},
  number={1},
  pages={92},
  year={2019},
  publisher={Nature Publishing Group UK London}
}

@article{balthasar2017metasurface,
  title={Metasurface polarization optics: independent phase control of arbitrary orthogonal states of polarization},
  author={Balthasar Mueller, JP and Rubin, Noah A and Devlin, Robert C and Groever, Benedikt and Capasso, Federico},
  journal={Physical review letters},
  volume={118},
  number={11},
  pages={113901},
  year={2017},
  publisher={APS}
}

@article{hsiao2017fundamentals,
  title={Fundamentals and applications of metasurfaces},
  author={Hsiao, Hui-Hsin and Chu, Cheng Hung and Tsai, Din Ping},
  journal={Small Methods},
  volume={1},
  number={4},
  pages={1600064},
  year={2017},
  publisher={Wiley Online Library}
}

@article{wang2017broadband,
  title={Broadband achromatic optical metasurface devices},
  author={Wang, Shuming and Wu, Pin Chieh and Su, Vin-Cent and Lai, Yi-Chieh and Hung Chu, Cheng and Chen, Jia-Wern and Lu, Shen-Hung and Chen, Ji and Xu, Beibei and Kuan, Chieh-Hsiung and others},
  journal={Nature communications},
  volume={8},
  number={1},
  pages={187},
  year={2017},
  publisher={Nature Publishing Group UK London}
}

@article{yang2020all,
  title={All-dielectric metasurface for high-performance structural color},
  author={Yang, Wenhong and Xiao, Shumin and Song, Qinghai and Liu, Yilin and Wu, Yunkai and Wang, Shuai and Yu, Jie and Han, Jiecai and Tsai, Din-Ping},
  journal={Nature communications},
  volume={11},
  number={1},
  pages={1864},
  year={2020},
  publisher={Nature Publishing Group UK London}
}

@article{arbabi2017planar,
  title={Planar metasurface retroreflector},
  author={Arbabi, Amir and Arbabi, Ehsan and Horie, Yu and Kamali, Seyedeh Mahsa and Faraon, Andrei},
  journal={Nature Photonics},
  volume={11},
  number={7},
  pages={415--420},
  year={2017},
  publisher={Nature Publishing Group UK London}
}

@article{faraji2018compact,
  title={Compact folded metasurface spectrometer},
  author={Faraji-Dana, MohammadSadegh and Arbabi, Ehsan and Arbabi, Amir and Kamali, Seyedeh Mahsa and Kwon, Hyounghan and Faraon, Andrei},
  journal={Nature communications},
  volume={9},
  number={1},
  pages={4196},
  year={2018},
  publisher={Nature Publishing Group UK London}
}

@article{kim2024metasurface,
  title={Metasurface folded lens system for ultrathin cameras},
  author={Kim, Youngjin and Choi, Taewon and Lee, Gun-Yeal and Kim, Changhyun and Bang, Junseo and Jang, Junhyeok and Jeong, Yoonchan and Lee, Byoungho},
  journal={Science Advances},
  volume={10},
  number={44},
  pages={eadr2319},
  year={2024},
  publisher={American Association for the Advancement of Science}
}

@article{mengu2020misalignment,
  title={Misalignment resilient diffractive optical networks},
  author={Mengu, Deniz and Zhao, Yifan and Yardimci, Nezih T and Rivenson, Yair and Jarrahi, Mona and Ozcan, Aydogan},
  journal={Nanophotonics},
  volume={9},
  number={13},
  pages={4207--4219},
  year={2020},
  publisher={De Gruyter}
}

@article{zhou2020situ,
  title={In situ optical backpropagation training of diffractive optical neural networks},
  author={Zhou, Tiankuang and Fang, Lu and Yan, Tao and Wu, Jiamin and Li, Yipeng and Fan, Jingtao and Wu, Huaqiang and Lin, Xing and Dai, Qionghai},
  journal={Photonics Research},
  volume={8},
  number={6},
  pages={940--953},
  year={2020},
  publisher={Chinese Laser Press and Optical Society of America}
}

@inproceedings{real2019regularized,
  title={Regularized evolution for image classifier architecture search},
  author={Real, Esteban and Aggarwal, Alok and Huang, Yanping and Le, Quoc V},
  booktitle={Proceedings of the aaai conference on artificial intelligence},
  volume={33},
  pages={4780--4789},
  year={2019}
}

@inproceedings{cubuk2019autoaugment,
  title={Autoaugment: Learning augmentation strategies from data},
  author={Cubuk, Ekin D and Zoph, Barret and Mane, Dandelion and Vasudevan, Vijay and Le, Quoc V},
  booktitle={2019 IEEE/CVF conference on computer vision and pattern recognition (CVPR)},
  pages={113--123},
  year={2019},
  organization={IEEE}
}

@article{mosk2012controlling,
  title={Controlling waves in space and time for imaging and focusing in complex media},
  author={Mosk, Allard P and Lagendijk, Ad and Lerosey, Geoffroy and Fink, Mathias},
  journal={Nature photonics},
  volume={6},
  number={5},
  pages={283--292},
  year={2012},
  publisher={Nature Publishing Group UK London}
}

@article{zheng2015compact,
  title={Compact deep convolutional neural networks for image classification},
  author={Zheng, Zejia and Li, Zhu and Nagar, Abhishek and Kang, Woosung},
  journal={Proc. ICMEW},
  pages={1--6},
  year={2015}
}

@inproceedings{hu2018squeeze,
  title={Squeeze-and-excitation networks},
  author={Hu, Jie and Shen, Li and Sun, Gang},
  booktitle={Proceedings of the IEEE conference on computer vision and pattern recognition},
  pages={7132--7141},
  year={2018}
}

@inproceedings{ishii20102000,
  title={2000 fps real-time vision system with high-frame-rate video recording},
  author={Ishii, Idaku and Tatebe, Tetsuro and Gu, Qingyi and Moriue, Yuta and Takaki, Takeshi and Tajima, Kenji},
  booktitle={2010 IEEE International Conference on Robotics and Automation},
  pages={1536--1541},
  year={2010},
  organization={IEEE}
}

@article{pivnenko2021sub,
  title={Sub-millisecond switching of multi-level liquid crystal on silicon spatial light modulators for increased information bandwidth},
  author={Pivnenko, Mike and Li, Kun and Chu, Daping},
  journal={Optics Express},
  volume={29},
  number={16},
  pages={24614--24628},
  year={2021},
  publisher={Optical Society of America}
}

@article{zheng2023dual,
  title={Dual adaptive training of photonic neural networks},
  author={Zheng, Ziyang and Duan, Zhengyang and Chen, Hang and Yang, Rui and Gao, Sheng and Zhang, Haiou and Xiong, Hongkai and Lin, Xing},
  journal={Nature Machine Intelligence},
  volume={5},
  number={10},
  pages={1119--1129},
  year={2023},
  publisher={Nature Publishing Group UK London}
}

@article{goi2022direct,
  title={Direct retrieval of Zernike-based pupil functions using integrated diffractive deep neural networks},
  author={Goi, Elena and Schoenhardt, Steffen and Gu, Min},
  journal={Nature Communications},
  volume={13},
  number={1},
  pages={7531},
  year={2022},
  publisher={Nature Publishing Group UK London}
}

@article{luo2024meta,
  title={Meta-Optics Based Parallel Convolutional Processing for Neural Network Accelerator},
  author={Luo, Mingcheng and Xu, Tengji and Xiao, Shuqi and Tsang, Hon Ki and Shu, Chester and Huang, Chaoran},
  journal={Laser \& Photonics Reviews},
  volume={18},
  number={11},
  pages={2300984},
  year={2024},
  publisher={Wiley Online Library}
}

@article{huang2006extreme,
  title={Extreme learning machine: theory and applications},
  author={Huang, Guang-Bin and Zhu, Qin-Yu and Siew, Chee-Kheong},
  journal={Neurocomputing},
  volume={70},
  number={1-3},
  pages={489--501},
  year={2006},
  publisher={Elsevier}
}

@article{huang2015trends,
  title={Trends in extreme learning machines: A review},
  author={Huang, Gao and Huang, Guang-Bin and Song, Shiji and You, Keyou},
  journal={Neural Networks},
  volume={61},
  pages={32--48},
  year={2015},
  publisher={Elsevier}
}

@article{pierangeli2021photonic,
  title={Photonic extreme learning machine by free-space optical propagation},
  author={Pierangeli, Davide and Marcucci, Giulia and Conti, Claudio},
  journal={Photonics Research},
  volume={9},
  number={8},
  pages={1446--1454},
  year={2021},
  publisher={Chinese Laser Press and Optical Society of America}
}

@article{teugin2021scalable,
  title={Scalable optical learning operator},
  author={Te{\u{g}}in, U{\u{g}}ur and Y{\i}ld{\i}r{\i}m, Mustafa and O{\u{g}}uz, {\.I}lker and Moser, Christophe and Psaltis, Demetri},
  journal={Nature Computational Science},
  volume={1},
  number={8},
  pages={542--549},
  year={2021},
  publisher={Nature Publishing Group US New York}
}

@article{tancik2020fourier,
  title={Fourier features let networks learn high frequency functions in low dimensional domains},
  author={Tancik, Matthew and Srinivasan, Pratul and Mildenhall, Ben and Fridovich-Keil, Sara and Raghavan, Nithin and Singhal, Utkarsh and Ramamoorthi, Ravi and Barron, Jonathan and Ng, Ren},
  journal={Advances in neural information processing systems},
  volume={33},
  pages={7537--7547},
  year={2020}
}

@article{kamali2017angle,
  title={Angle-multiplexed metasurfaces: encoding independent wavefronts in a single metasurface under different illumination angles},
  author={Kamali, Seyedeh Mahsa and Arbabi, Ehsan and Arbabi, Amir and Horie, Yu and Faraji-Dana, MohammadSadegh and Faraon, Andrei},
  journal={Physical Review X},
  volume={7},
  number={4},
  pages={041056},
  year={2017},
  publisher={APS}
}

@article{liu2018metasurface,
  title={Metasurface enabled wide-angle Fourier lens},
  author={Liu, Wenwei and Li, Zhancheng and Cheng, Hua and Tang, Chengchun and Li, Junjie and Zhang, Shuang and Chen, Shuqi and Tian, Jianguo},
  journal={Advanced Materials},
  volume={30},
  number={23},
  pages={1706368},
  year={2018},
  publisher={Wiley Online Library}
}

@article{vandermaaten08a,
  title={Visualizing Data using t-SNE},
  author={Van der Maaten, Laurens and Hinton, Geoffrey},
  journal={Journal of Machine Learning Research},
  volume={9},
  pages={2579--2605},
  year={2008}
}

@ARTICLE{9066950,
  author={Yu, Yongbin and Adu, Kwabena and Tashi, Nyima and Anokye, Patrick and Wang, Xiangxiang and Ayidzoe, Mighty Abra},
  journal={IEEE Access}, 
  title={RMAF: Relu-Memristor-Like Activation Function for Deep Learning}, 
  year={2020},
  volume={8},
  number={},
  pages={72727-72741}}

@article{Zhu2021ImageRT,
  title={Image reconstruction through a multimode fiber with a simple neural network architecture},
  author={Y. Zhu and C. Ma and C. Y. Lin and Y. S. Lin and P. J. Tsai and Y. H. Lin and T. B. Tang and M. T. Lin and K. M. Lee and S. J. Tang},
  journal={Scientific Reports},
  volume={11},
  pages={896},
  year={2021}
}

@article{Agarwal2017FacialKP,
  title={Facial Key Points Detection using Deep Convolutional Neural Network - NaimishNet},
  author={Naimish Agarwal and Alexander Krohn-Grimberghe and Rishabh Vyas},
  journal={arXiv preprint arXiv:1710.00977},
  year={2017}
}

@article{juliano2022metasurface,
  title={Metasurface-enhanced light detection and ranging technology},
  author={Juliano Martins, Renato and Marinov, Emil and Youssef, M Aziz Ben and Kyrou, Christina and Joubert, Mathilde and Colmagro, Constance and G{\^a}t{\'e}, Valentin and Turbil, Colette and Coulon, Pierre-Marie and Turover, Daniel and others},
  journal={Nature communications},
  volume={13},
  number={1},
  pages={5724},
  year={2022},
  publisher={Nature Publishing Group UK London}
}

@article{gao2019highly,
  title={A highly efficient bifunctional dielectric metasurface enabling polarization-tuned focusing and deflection for visible light},
  author={Gao, Song and Park, Chul-Soon and Lee, Sang-Shin and Choi, Duk-Yong},
  journal={Advanced Optical Materials},
  volume={7},
  number={9},
  pages={1801337},
  year={2019},
  publisher={Wiley Online Library}
}
\bibliographystyle{sciencemag}

% After the paper has completed peer review and been revised ready for acceptance,
% you should comment out the lines above and copy-paste the contents of your .bbl
% file here instead. This will help ensure that our conversion software works correctly.
% Remember to re-run BibTeX first - check the timestamp!
%
% Example of the first three entries copy-pasted from science_template.bbl:
%
%\begin{thebibliography}{1}
%
%\bibitem{example}
%A.~N. {Author}, An example reference. \emph{Journal of Improbable Research}
%  \textbf{1}, 67 (2020).
%
%\bibitem{example2}
%F.~M. {Surname}, S.~{Author}, A second example. \emph{Interesting Research
%  Letters} \textbf{32}, 897 (2019).
%
%\bibitem{example_preprint}
%P.~{One}, P.~{Two}, P.~{Three}, {An unpublished preprint}. \emph{preprint}
%  (2021), arXiv:2101.12345.
%
%\end{thebibliography}

%%%%%%%%%%%%%%%% ACKNOWLEDGEMENTS %%%%%%%%%%%%%%%

\section*{Acknowledgments}

\paragraph*{Funding:}
This work was supported by RGC GRF 14208925, ECS 24203724, YCRG C4004-24Y, YCRG C1002-22Y, NSFC 62405258, ITF ITS/237/22, NSFC/RGC N\_CUHK444/22.
\paragraph*{Author contributions:}
C.H. and M.L. conceived the idea. M.L. designed the metasurface chip and conducted the experiments. M.L. and J.X. conducted simulations. M.L.and C.H. analyzed the experimental data. C.H. and M.L. led the writing of the main manuscript. C.S., J.P., J.H., and H.C. contributed to the paper writing. C.H. and C.S. supervised the project. 
\paragraph*{Competing interests:}
There are no competing interests to declare.
\paragraph*{Data, code and materials availability:}
The data that support the plots within this paper and other findings of this study are available from the corresponding author upon reasonable request.
\paragraph*{Supplementary information:}
Supplementary material is available at the submission materials.
%%%%%%%%%%%%%%%% SUPPLEMENT LIST %%%%%%%%%%%%%%%

% List the contents of your Supplementary Materials, including the numbers of any
% supplementary figures, tables, external data files etc. and any references that are
% cited only in the supplement. In this example, refs. 7-8 are cited only in the supplement.
% Fill out your numbers accordingly and delete any lines that aren't applicable.

%%%%%%%%%%%%%%%% SUPPLEMENT TITLE PAGE %%%%%%%%%%%%%%%

\end{document}